\documentclass[11pt]{article}
\usepackage[margin=2.5cm]{geometry}
\usepackage{float}
\usepackage{times}
\usepackage{enumitem}
\usepackage{titlesec}
\usepackage[backend=biber, style=authoryear-ibid]{biblatex}
\usepackage[T1]{fontenc}
\usepackage{array}
\usepackage{hyperref}
\usepackage[nameinlink]{cleveref}
\usepackage{xurl}
\hypersetup{breaklinks=true}

\usepackage{tocloft}
\renewcommand{\cftsecfont}{\bfseries} 

\renewcommand{\thesection}{\arabic{section}}
\renewcommand{\thesubsection}{\arabic{section}.\Alph{subsection}}

\cftsetindents{section}{0em}{2.5em}
\cftsetindents{subsection}{1.5em}{3em}

\usepackage{titlesec}
\titleformat{\section}
  {\huge}{\thesection}{2em}{}
\usepackage{titlesec}
\titleformat{\subsection}
  {\Large}{\thesubsection}{1em}{}

\renewcommand\thesection{\arabic{section}}
\renewcommand{\thesubsection}{\thesection.\Alph{subsection}}
\renewcommand*{\bibfont}{\normalfont\small}

\crefformat{section}{\textbf{#2Section~#1#3}}
\Crefformat{section}{\textbf{#2Section~#1#3}}
\Crefformat{appendix}{\textbf{#2Appendix~#1#3}}

\crefmultiformat{section}{\textbf{#2Sections~#1#3}}{ and \textbf{#2#1#3}}{, \textbf{#2#1#3}}{, and \textbf{#2#1#3}}
\Crefmultiformat{section}{\textbf{#2Sections~#1#3}}{ and \textbf{#2#1#3}}{, \textbf{#2#1#3}}{, and \textbf{#2#1#3}}

\crefmultiformat{appendix}{\textbf{#2Appendices~#1#3}}{ and \textbf{#2#1#3}}{, \textbf{#2#1#3}}{, and \textbf{#2#1#3}}
\Crefmultiformat{appendix}{\textbf{#2Appendices~#1#3}}{ and \textbf{#2#1#3}}{, \textbf{#2#1#3}}{, and \textbf{#2#1#3}}
\begin{document}

\begin{titlepage}

\title{\Huge The GPT Dilemma: Foundation Models and the Shadow of Dual-Use \\ \vspace{1cm} \Large Navigating the Ambiguities of Civilian and Military Applications in the AI Era }

\author{\textbf{Alan Hickey} \\ OpenAI \\ alan@openai.com}

\date{July, 2024}
\maketitle

\begin{abstract}\noindent \small This paper examines the dual-use challenges of foundation models and the consequent risks they pose for international security. As artificial intelligence (AI) models are increasingly tested and deployed across both civilian and military sectors, distinguishing between these uses becomes more complex, potentially leading to misunderstandings and unintended escalations among states. The broad capabilities of foundation models lower the cost of repurposing civilian models for military uses, making it difficult to discern another state’s intentions behind developing and deploying these models. As military capabilities are increasingly augmented by AI, this discernment is crucial in evaluating the extent to which a state poses a military threat. Consequently, the ability to distinguish between military and civilian applications of these models is key to averting potential military escalations. The paper analyzes this issue through four critical factors in the development cycle of foundation models: model inputs, capabilities, system use cases, and system deployment. This framework helps elucidate the points at which ambiguity between civilian and military applications may arise, leading to potential misperceptions. Using the Intermediate-Range Nuclear Forces (INF) Treaty as a case study, this paper proposes several strategies to mitigate the associated risks. These include establishing red lines for military competition, enhancing information-sharing protocols, employing foundation models to promote international transparency, and imposing constraints on specific weapon platforms. By managing dual-use risks effectively, these strategies aim to minimize potential escalations and address the trade-offs accompanying increasingly general AI models.

\end{abstract}

\end{titlepage}

\tableofcontents

\clearpage
\section{Introduction}

As foundation models continue to advance in their capabilities, it is crucial for researchers, policymakers, and developers to understand their potential impact on international stability. This article makes several contributions. It identifies the importance of foundation models in international stability and develops a framework to assess the risks. This framework examines the conditions under which the uses of foundation models are distinguishable between civilian and military applications. It focuses on the role of foundation models in reducing distinguishability, because of their ability to carry out a broad range of tasks and their potential to be deployed in a variety of settings where an adversary’s intentions are unclear. This paper also suggests a number of paths forward in reducing risks created by the development and adoption of these models. This should serve as a basis for examining the appropriate contexts and methods for deploying foundation models in society.

I use the term foundation models to refer to large-scale machine learning models that serve as a basis for various downstream tasks. They are pre-trained on vast datasets and can be fine-tuned for specific applications. These models are very powerful, showing abilities to assist with writing articles \autocite{openai_gpt-4_2024}, generating code \autocite{openai_gpt-4_2024}, assessing data \autocite{morgan_exploring_2023, nejjar_llms_2024}, conducting legal analysis \autocite{savelka_unlocking_2023}, and assisting in medical diagnoses \autocite{zhang_challenges_2024}. The diverse abilities of foundation models are evident in their application across different modalities, including visual synthesis with models like DALL-E 3 and Midjourney \autocite{openai_dalle_2023, midjourney_midjourney_nodate}, music composition with AudioCraft and speech synthesis with NaturalSpeech 2 \autocite{meta_ai_audiocraft_nodate, shen_naturalspeech_2023}.

Foundation models are emerging as general-purpose technologies because of their ability to carry out a broad range of tasks, including many tasks which humans carry out \autocite{eloundou_gpts_2023}. Eloundou et al. have found that applications built on foundation models could help carry out around half of the tasks in the economy \autocite{zhang_one_2023}. Indeed, foundation models are being adopted across a broad range of industries and are continuing to diffuse across the economy \autocite{gupta_empirical_2024}. They also have significant scientific potential, with AlphaFold predicting structure of proteins and assisting in drug discovery \autocite{jumper_highly_2021, chakraborty_artificial_2023}. Foundation models have also demonstrated abilities to generalize beyond their original training. GPT-4, for instance, was able to navigate a map and draw a picture with LaTeX, despite not being trained for either of these tasks \autocite{bubeck_sparks_2023}.

Capabilities are likely to improve for a number of reasons. Scaling laws mean that model capabilities improve with increased training compute \autocite{hestness_deep_2017, kaplan_scaling_2020, villalobos_scaling_2023}. Among frontier models, which are the most advanced models, the trend 2015–2022 was of doubling the amount of training compute approximately every ten months \autocite{openai_frontier_2023, sevilla_compute_2022}. Orders of magnitude more compute are likely to be added in the future, as companies continue advancing performance. Quantifying algorithmic improvements is more complicated because of the lack of a single performance metric \autocite{hernandez_measuring_2020}. In addition, it is difficult to quantify changes in the types of problems which foundation models can solve, although trends indicate significant algorithmic improvements over time \autocite{dahl_benchmarking_2023}. Since we are running out of additional training data from the internet, synthetic data produced by models will likely have to fill the gap \autocite{kaplan_scaling_2020}. The results of this are unclear, though early results appear promising \autocite{lu_machine_2024}.

The adoption of general-purpose technologies across a state’s economy tends to be associated with accumulating power in the international system \autocite{modelski_leading_1996, ding_rise_2024}. The economic development which follows gives states the capacity to develop their military capacity, exert influence abroad and build international coalitions \autocite{beckley_power_2018}. States possessing the organizational capacity to harness these technologies effectively will translate these economic gains into military innovations. Successful integration of general-purpose technologies into the military sector can enhance a state's power projection and warfare capabilities \autocite{horowitz_diffusion_2010}. The introduction of electricity in naval ships led to ships that ran faster, had improved communication capabilities and could more successfully evade detection than non-electrified ships \autocite{ding_engines_2023}.

AI has many potential military applications, some of which have already been deployed. These include autonomous weapons systems (AWS) \autocite{us_department_of_defense_autonomy_2023}, target recognition \autocite{boury-brisset_benefits_2020}, and predictive maintenance and training \autocite{hannas_chinas_2022}. Some existing work related to deterrence and AI focuses on the effects of specific weapons systems, like AWS. Sterbenz and Trager argue that AWS decrease the costs of force deployment and mobilization, since they reduce the need for soldiers on the ground \autocite{sterbenz_autonomous_2019}. While this initially reduces the human death toll, it may increase the likelihood of escalation. Horowitz emphasizes the potential escalation risks tied to the increased speed of AI-facilitated decision-making, and the lack of sufficient room for an off-ramp \autocite{horowitz_when_2019}. Escalation therefore might become more rapid with the introduction of autonomous weapons systems \autocite{sterbenz_autonomous_2019}.

Many states agree on a set of principles to guide the development of AI in a military context \autocite{us_department_of_state_political_nodate}, though it remains to be seen what this means in practice for testing and evaluation. The US Department of Defense’s Directive 3000.09 outlines the need for the operator continuing to exercise “appropriate levels of human judgment” \autocite{sayler_defense_2024}. Major powers, however, disagree on the definitions and approaches to autonomous weapons \autocite{taddeo_comparative_2022}. Meanwhile, there is a growing concern from civil society over the reduced role of human decision-making as AI technologies advance and become more integrated into military operations \autocite{klare_address_2023}.

Foundation models are distinct from traditional machine learning because of their generalizability—their ability to carry out useful tasks with low costs of switching to another domain. As states experiment with different use cases, they are demonstrating a variety of potential uses in military and diplomatic settings. These include disinformation campaigns \autocite{goldstein_generative_2023}, threat monitoring and decision support \autocite{vynck_tech_2023}, target recognition \autocite{tarantola_palantir_2023}, intelligence, surveillance and reconnaissance (ISR) \autocite{hannas_chinas_2022}, supporting command and control \autocite{saltini_ai_2023}, facilitating offensive cyber operations \autocite{beauchamp-mustafaga_exploring_2024}, military planning and strategy \autocite{jensen_how_2023}, and research and development \autocite{darpa_ai_nodate}. Whether and how states integrate foundation models into military systems will thus have widespread implications for how states maintain stability and engage in conflict. Despite this, there is surprisingly little work on general-purpose technologies and deterrence.

Foundation models are likely to have direct effects on state power as they are integrated within military systems, acting as force multipliers for both existing and future weapons platforms and supporting infrastructure. They will also have indirect effects as models permeate throughout the economy, boosting productivity, trade, and scientific research. While some capabilities may be purely military or civilian, many capabilities sit in between, like models used to assist in research and development.

The concepts of dual-use and distinguishability are central to discussions of foundation models. Distinguishability refers to how easily other states can differentiate between civilian and military uses of a technology \autocite{vaynman_dual_2023}. As general-purpose technologies, foundation models significantly lower the cost of switching domains, meaning that a model designed for civilian purposes could more easily be repurposed for military uses than narrow ML models. In this article, I argue that a primary challenge in international relations, in the era of advanced AI, is navigating the uncertainties surrounding an adversary's development, deployment, or utilization of such models. Effectively managing competition among technologically advanced states will hinge on their capacity to differentiate between the diverse applications of these models, particularly in distinguishing civilian applications from military ones.

\section{The GPT Dilemma}

The core challenge of foundation models is what we can call the GPT dilemma: foundation models will enable significant productivity advances in the civilian sector, while simultaneously bolstering military capabilities. As a result, states will be motivated to incorporate them throughout their economies. In this context, determining whether another state poses a military threat will necessitate discerning the intentions behind its use of foundation models. Thus, as states compete over foundation models, the ability to distinguish between military and civilian applications of these models will be crucial for preventing military escalations. This paper will investigate the conditions under which civilian and military uses of foundation models will be distinguishable.

Vaynman and Volpe identify two important characteristics of dual-use technologies—those which have both civilian and military applications—namely their distinguishability and integration \autocite{vaynman_dual_2023}. Low levels of distinguishability increase the amount of information required for an adversary to differentiate between civilian and military use cases. In cases of low distinguishability, such as space or cyber technology, other states are more uncertain of the purpose of a deployment. Related to this, integration is defined by the breadth and depth of incorporation of the technology in both civilian and military sectors. When technologies are highly integrated, like drones, it increases the likelihood of adversary inspections revealing military vulnerabilities. Vaynman and Volpe argue that these characteristics help determine the extent to which a technology can be monitored, with highly integrated and low distinguishability technologies being the most difficult to monitor.

My analysis builds on this framework but concentrates on the dual-use capabilities of foundation models. These models are highly integrated technologies, because of their ability to combine with many other technologies and carry out a broad range of actions. Foundation models tend to have relatively low distinguishability between their civilian and military applications, though the degree of distinguishability can vary significantly. This variability is partly due to the emerging nature of the technology, which means the full extent of distinguishability is yet to be determined, and partly because of the inherent characteristics of foundation models themselves.

In contrast to integration, which tends to be most relevant at the point of deployment, distinguishability tends to vary significantly across the development cycle. Thus, I extend Vaynman and Volpe’s framework by assessing how the incentives for states to compete or cooperate vary at different points in a technology’s development. States, for instance, may cooperate in tracking fissile material to prevent nuclear proliferation but compete over the deployment of nuclear weapons. However, since the key stages and processes for developing a technology vary greatly, I will focus specifically on foundation models and not attempt to generalize this framework to other technologies. Whereas Vaynman and Volpe focus on the similarities between different technologies, I concentrate on the unique characteristics of foundation models.

To systematically assess distinguishability, I look at the four main factors affecting development and deployment: model inputs, model capabilities, system use cases, and system deployment. I will briefly give an overview of these factors, before assessing each in the following subsections. I argue that dual-use problems arise at every stage in the deployment cycle and that distinguishability is a particularly useful framework to assess this in the context of foundation models.

The first factor is model inputs. In the case of foundation models, data, algorithms, and computing power (compute) are most relevant \autocite{scharre_four_2024}. Though other inputs are necessary, like energy and labor, these are generally paired with one of these three inputs. Energy, for instance, is necessary to power compute. Labor is necessary for all three of the inputs. Since data, algorithms, and compute are all broadly useful inputs, establishing civilian-military distinguishability is difficult at this stage, though compute offers the most promising approach. This dual-use challenge is evident because the same data, algorithms, and compute can serve both civilian and military objectives in foundation models. For instance, compute allocated for climate modeling can equally support simulations for strategic military planning. The inherent versatility and broad applicability of these inputs make it challenging to differentiate between civilian and military intentions at this early stage.

The second factor is a model’s capabilities. After training, the model has a set of encoded knowledge. I will argue that foundation models with greater generalizability and domain knowledge have reduced distinguishability. The generality of foundation models means they exhibit substantial civilian and military applications, driving down convertibility costs. This will make it challenging for states to signal purely civilian intentions in adopting this technology \autocite{volpe_dual-use_2019}. The dual-use nature of model capabilities is particularly pronounced because these models can be adapted for a wide range of tasks. For example, a foundation model trained to assist in medical diagnostics could also be adapted for battlefield triage or decision-making in combat scenarios. A language model used for customer support can be repurposed for intelligence analysis in military contexts. The ability of these models to transition between different applications highlights the difficulty in distinguishing between civilian and military uses. Thus, the problem of foundation models in international security will be one of establishing grounds for cooperation despite the challenges which dual-use capabilities present. The uncertainty between states over the capabilities of their adversary's foundation models, and what their capabilities might be in different settings or if fine-tuned on specific data, further complicates efforts at establishing clear distinguishability between civilian and military applications.

System use cases are the third factor to consider regarding foundation model distinguishability. After a foundation model is trained, it is often fine-tuned, tested, and deployed through a system. Users interact with foundation models through applications such as ChatGPT or Claude. Third-party developers use APIs to build applications on top of the models, and developers can also chain tools like Wolfram|Alpha to the model to combine it with more specialized applications.

The dual-use nature arises here as well. Civilian applications, such as customer service chatbots, can be repurposed for military communications or psychological operations. Systems like Donovan and AIP are being tested for uses in battlefield assessment and planning \autocite{scale_ai_donovan_nodate, palantir_aip_nodate}. Fine-tuning a foundation model for commercial logistics optimization can similarly be adapted for military supply chain management. In addition to software components, foundation models have the potential to be paired with hardware. For example, future systems could be deployed in a command-and-control center to manage AWS. The versatility of these systems in various contexts highlights the challenge in distinguishing between civilian and military deployments. The ability to reconfigure and adapt these models for different purposes further blurs the lines, complicating efforts to establish clear boundaries between civilian and military use cases.

I also consider the information environment as part of this third factor. Foundation models will change civilian-military distinguishability through the relative costs of generating and analyzing information. Many researchers have argued that foundation models will continue decreasing the cost of generating information \autocite{goldstein_generative_2023}. This declining cost, some argue, will make it more challenging to ascertain accurate information due to an increase in data poisoning and misinformation. Such effects could manifest across a range of domains. For instance, it could include changing the mind of some target audience of an adversary’s population or reducing confidence among military commanders that they are receiving accurate information about the location of their adversary’s missiles. However, beyond generating information, foundation models can also assess information at scale. By identifying relevant data and assessing patterns across a range of domains, foundation models could make it easier to distinguish between civilian and military activity. The net effect of foundation models in this area will depend on whether their improved analytical abilities outweigh their contribution to decreasing the quality of the information environment.

Finally, I consider deployment as the final factor. The role of dual-use is affected by how humans interact with these systems and how institutions manage them. Government regulations can play a significant role by mandating specific rules around defense deployment standards. Civilian-military convertibility costs refer to how difficult it is to convert civilian capabilities into military capabilities. This includes costs such as research and development for specific military applications, combining foundation models with relevant military hardware, testing additional use cases in a military context, and ensuring compatibility with existing defense systems. This notion of convertibility is established in the literature and encompasses scenarios where researchers deliberately seek to convert civilian capabilities into military ones, as well as cases where this conversion occurs unintentionally \autocite{cao_conversion_2020}. It also incorporates the speed of conversion—the longer it takes to convert from civilian to military use, the greater the ability to distinguish between the two. The concept of convertibility is crucial for dual-use and distinguishability, as higher convertibility costs imply greater difficulty in repurposing civilian technologies for military use, making it easier to distinguish between civilian and military deployments. Deployment policies within the industry significantly affect distinguishability because they influence transparency and set norms for user interactions with these systems. For example, platform policies determine how users can access and use the models, and how violations are detected and handled.

Distinguishability is also affected by the extent to which civilian capabilities spill over into military capabilities. The partners that companies choose to work with, such as the Department of Defense, and how such deployments are implemented affect transparency and perceived intent. Establishing clear policies and standards for the deployment of foundation models in both civilian and military contexts would help to clarify their uses and mitigate dual-use risks. These deployment policies directly impact dual-use by determining which users can access the product and under what conditions, thereby influencing how easily civilian technologies can be repurposed for military use.

It is worth noting that the four factors affecting the development and deployment of foundation models are not clear and discrete stages, and they often happen in parallel. For instance, models will often continue to be fine-tuned after the system has been deployed. Dividing these components of development and deployment remains useful for assessing distinguishability at different points.

This framework offers several key advantages. First, it emphasizes dual-use considerations. In a technology that is being deployed so widely, cooperation between states will depend on managing the dual-use nature of foundation models. Second, it addresses trade-offs at every stage of development and deployment. These models can be repurposed for both civilian and military applications at every stage, from research and development to deployment. Third, it accounts for perception ambiguity, acknowledging that the way foundation model development and deployment is perceived can vary greatly depending on context and intent.

\begin{tabular}[h]{|m{0.12\linewidth}|>{\raggedright\arraybackslash}m{0.18\linewidth}|>{\raggedright\arraybackslash}m{0.2\linewidth}|>{\raggedright\arraybackslash}m{0.4\linewidth}|}
    \hline
    \textbf{Factor} & \textbf{Components} & \textbf{Distinguishability} & \textbf{Policies} \\ \hline
    & Data & Low & Information sharing between AI labs and between governments \\ \cline{2-4}
    \centering Inputs & Algorithms & Very low & Information sharing between AI labs and between governments \\ \cline{2-4}
    & Compute & Low/Medium & Tracking compute; establish monitoring systems; differential hardware development paths \\ \hline
    & Generalizability & Decreases as models generalize & Red-teaming and evaluations; implement transparent assessment protocols \\ \cline{2-4}
    \centering Model & Domain Knowledge & Decreases as capabilities improve in key domains & Develop domain-specific evaluations; particularly for military use cases \\ \cline{2-4}
    & Capability Uncertainty & Decreases if there is greater uncertainty & Enhance transparency in capability development; share less sensitive defense evaluation results \\ \hline
    & Hardware & High & Conduct joint military-civilian assessments; establish clearer testing and evaluation protocols for defense use cases; constraining weapons platforms \\ \cline{2-4}
    \centering System & Software & Medium & Conduct joint military-civilian assessments; robust platform abuse monitoring and enforcement \\ \cline{2-4}
    & Information Environment & Medium & Deploy models to assist intelligence and military analysts; AI interpretability research \\ \hline
    \centering Deployment & Specialization & Increases if civilian and military deployments are more specialized & Establish red lines for military competition; share testing, evaluation and deployment practices - particularly among allies; establish clearer deployment protocols \\ \cline{2-4}
    & Capability spillover & Decreases if civilian capabilities spill over into military capabilities & Establish clearer regulations for military testing and deployment for AI; AI labs establish rules for military cooperation \\ \hline
\end{tabular}

Distinguishability is often framed in terms of the security dilemma. In brief, the security dilemma exists because of  international anarchy. States must provide for their own security, but in an environment of imperfect information, defensive capabilities may be misinterpreted as offensive, thus raising the potential for conflict. The extent of the security dilemma depends on the offense-defense balance and offense-defense distinguishability \autocite{jervis_cooperation_1978}. The former describes whether it is more costly to gain territory or hold existing territory and the latter describes the degree to which a state can discern between the offensive and defensive capabilities of another state. In scenarios where the ability to make this distinction is reduced, the odds of states spiraling into conflict increase. 

A significant limitation of the offense-defense framing of distinguishability is that dual-use problems arise at every stage in the development and deployment of foundation models.\footnote{The offense-defense approach has faced criticism due to the vagueness of its claims, challenges in operationalizing its variables, and questions about its predictive power: \autocite{levy_offensivedefensive_1984}} Compute is a key input for producing foundation models. How should a state interpret another state's computing resources—as offensive or defensive? Defining what constitutes offensive and defensive capabilities with respect to model inputs is conceptually challenging. The ability to distinguish between civilian and military capabilities precedes offense-defense distinguishability \autocite{volpe_dual-use_2019}. Before determining whether a state's capabilities are being used for offensive or defensive purposes, one must first establish whether these capabilities are being used for military or civilian purposes \autocite{lupovici_dual-use_2021}. Civilian-military distinguishability encompasses each stage of development, such as research, testing, and deployment, whereas offense-defense does not apply at every stage. Addressing the problem of dual-use capabilities also offers a more plausible approach, as civilian and military capabilities are more readily distinguishable, despite the challenges.

Schneider offers an alternative perspective, suggesting that military revolutions are fueled by the advancement of new technologies that necessitate resource utilization \autocite{schneider_capabilityvulnerability_2019}. This dependence on resources introduces vulnerabilities that adversaries of a state might exploit, thereby leading to instability. Though there is undoubtedly utility to the capability-vulnerability paradox, the core problem of foundation models extends beyond resource-induced vulnerabilities to encompass the challenge of their dual-use capabilities. To be sure, access to energy and compute will be highly important facets of international competition in the age foundation models. But this constitutes just one stage in how foundation models are developed and deployed. It would be difficult to apply the capability-vulnerability paradox beyond assessing a model’s inputs. Applying the capability-vulnerability paradox therefore does not fully encompass the broader competitive dynamics which foundation models introduce.

Instead, the role foundation models will play in international security is best characterized as a type of information problem, where states lack adequate accurate information on the intentions of other states. Their effects will be on how they combine with other technologies and how they re-shape the information environment. The challenge for external actors lies in inferring the intentions of states throughout the development and integration phases of these models. In the next subsections of this piece, I will examine each of these stages. After this, I present a case study and discuss promising approaches to addressing the problem of distinguishability.

\subsection{Model Inputs}

Model inputs provide states with the latent capacity to produce foundation models. The inputs to a model are generally the least distinguishable, but also the least militarily threatening. These inputs --- compute, data and algorithms --- are necessary to train a foundation model \autocite{scharre_four_2024}. I argue that each of these components pose significant challenges for distinguishability, though compute offers some promise.

The training process comprises two primary stages: pre-training and fine-tuning. During pre-training, the goal is to instill broad knowledge into the model. It is exposed to vast datasets, including from the internet, and is trained to recognize patterns within this data \autocite{openai_how_2023}. For instance, a language model might be trained to predict the subsequent word in a sentence. Through such exercises, the model discerns and internalizes general patterns and structures present in the training data. After this, the model goes through fine-tuning to improve its general usefulness and, in some cases also to improve its abilities in specific domains. To achieve the latter, more concentrated, task-specific datasets are utilized. For example, if the objective is to support ISR, the model could be fine-tuned using pertinent satellite and sensor data. A model may have general capabilities in object recognition after pre-training, while fine-tuning could result in it being able to detect salient movements of an adversary’s missiles. 

In the context of compute, several key aspects need to be assessed: tracking compute, identifying if the compute is being used for foundation models, discerning between its use for training or inference, and determining whether it is intended for civilian or military purposes.

The tracking of compute, unlike the more established methodologies in the nuclear sector, is not particularly well-developed. However, leveraging approaches from the nuclear industry has suggested viable methods for tracking compute resources. In the nuclear field, states developing capabilities often undergo inspections, with a focus on the traceability of uranium, which is essential for nuclear development. These materials leave a detectable footprint, aiding inspectors while still preserving state secrets. Applying similar principles to compute tracking, we can focus on the equipment used in semiconductor manufacturing, particularly Extreme Ultraviolet (EUV) lithography machines. Currently, only ASML can produce EUV machines. ASML has a small number of buyers and EUV machines are subject to export controls, making them relatively traceable. Future methods could involve a system that combines facility inspections with satellite monitoring to track EUV machines or large quantities of compute \autocite{baker_nuclear_2023}. As I will discuss in more detail later, the analytical capabilities of foundation models, paired with the declining cost of sensors, may offer some promise in verifying compliance with export controls, treaties or international law.

After tracking compute, states must ascertain whether this compute is being used for training foundation models or for running inference. Training frontier-level foundation models requires exceptionally high compute resources. The magnitude of these demands, such as intense power consumption, water requirements and heat generation, could indicate if compute is being used for training purposes through satellite imagery and sensor data. Thus, if a state acquires large amounts of leading-edge compute, along with hiring the necessary workers and other factors of production, this may signal the development of foundation models. 

Currently, inference for frontier models predominantly occurs in large data centers. We should soon expect the development of devices which conduct local inference. Researchers suggest that semiconductors optimized for inference will be distinguishable from those used for training \autocite{sastry_computing_2024}. Regardless, size and power constraints of local devices mean most inference will probably mostly stay at data centers.

States aiming to discern whether compute is being used for civilian or military purposes face a significant challenge because of the general-purpose nature of semiconductor chips. Unlike nuclear technology, which has clear delineations in its applications for weapons and power, chips have a wide array of uses in various sectors such as smartphones, gaming, and finance. This versatility provides states with many plausible reasons to develop or procure such technology. Though chips intended for civilian purposes could theoretically be repurposed for military systems, this is likely to be inefficient, as military applications typically have different specification requirements compared to civilian uses. For military uses, especially in operational theaters, attributes like durability, security, and reliability are often more crucial than speed. Existing AI models used by militaries tend to employ traditional machine learning with significantly lower compute requirements, tailored to the specific demands of military environments \autocite{hoffman_reducing_2023}. Even as militaries adopt foundation models, these systems will likely use mature nodes, rather than leading-edge chips. This distinction in requirements between civilian and military compute uses, despite the general-purpose nature of chips, provides some guidance for states trying to determine the end-use of these technologies.

However, distinguishability may be more difficult when it comes to use cases like data processing for ISR or the use of models for military R\&D, which could use some off-the-shelf chips. In this case, inferring intent is highly contextual and requires significant information about the actor using the chip. One way which US policymakers have sought to limit the proliferation of frontier capabilities to China is through export controls on chips 16nm or below.\footnote{Nanometer in this context is a marketing term, since it does not actually refer to the size of chips. But smaller nm chips signify more advanced chips \autocite{us_bis_commerce_2022}.} The training of frontier models requires chips below this threshold \autocite{sastry_computing_2024}. But it could be difficult to infer whether a state is developing such models for use in military systems. Hence, the regulation and monitoring of computing resources is a coarse tool, since its restrictions will often reduce both civilian and military capacities alike.

Fine-tuning requires significantly less data than pre-training. As a result, fine-tuning means capabilities can change at relatively low cost. This low cost makes distinguishability more difficult. The quality and the characteristics of the fine-tuning dataset is particularly important. If the dataset has biases, is poisoned, or is otherwise not representative of real-world use cases, this will lead to inaccurate outputs. 

The model’s data is distinguishable in certain contexts. For instance, you can improve capabilities in a given domain through fine-tuning. Inspecting the data used to fine-tune a model will reveal more information about the actor than inspecting the pre-training data, since the pre-training data will typically be more general and significantly larger. A model can be pre-trained for civilian purposes, but since fine-tuning is considerably cheaper than pre-training, it could be inexpensive to convert this for military purposes.

Algorithms are a critical and sensitive aspect of foundation models, with their broad principles widely understood but specific details kept as closely guarded secrets by those developing frontier models. Weights and biases, which constitute the model, are proprietary knowledge that provide a competitive edge. Notably, these parameters are stored in relatively small files. This compactness poses a risk as it simplifies the potential transfer or misappropriation of a model without easy detection. The complexity and proprietary nature of these components, coupled with their small file size, make algorithms an unlikely candidate for increasing model distinguishability.

\subsection{Model Capabilities}

As foundation models become increasingly capable, the ability to distinguish between civilian and military capabilities will likely diminish. I identify three factors pertaining to model capabilities which affect distinguishability—generalizability, domain knowledge and capability uncertainty.

By generalizability, in this context, I mean the ability to make reliable predictions, identify patterns, and extrapolate from the data in pre-training and post-training. The very nature of increasingly capable foundation models obscures the boundary between military and civilian uses. Since foundation models are becoming more general purpose, sorting and categorizing what is military and what is civilian is \textit{in principle} very difficult. This is a significant challenge for distinguishability. Some capabilities are more likely to blur the civilian-military distinction because of their usefulness in a variety of contexts \autocite{shevlane_model_2023}. For example, helping with military strategy requires some combination of strategic awareness, long-term planning, and domain-specific knowledge, yet strategic awareness and long-term planning would also be useful for a variety of enterprise applications. 

Beyond this, generalizability affects the model’s capacity to inflict harm. For instance, the narrow knowledge of AI systems that power self-driving cars limits their capacity for misuse and constrains the range of potential harm by limiting its capacity to a single domain. Conversely, systems endowed with broad knowledge can be useful in both civilian and military contexts. Thus, greater generalizability means being able to overcome a more heterogeneous set of barriers to conducting tasks. 

Domain capabilities refers to the ability of a model to achieve tasks in a given area, be it legal, medical, pedagogical, technical, or any other area. Improved capabilities in certain domains present greater risks. For instance, the October 30th, 2023, White House Executive Order on AI establishes rules for reporting the development of AI models, setting a lower compute threshold for those primarily trained on biological sequence data than those required for general-purpose models \autocite{biden_executive_2023}. This recognizes the fact that, beyond a certain threshold, abilities in the biological domain are particularly likely to have a high-risk profile, particularly when distinguishing between necessary research and risky activities may be difficult. 

Current models are very limited in their ability to inflict harm. These limitations are instructive of emerging risks. Though they can provide some information on CBRN, their knowledge is likely insufficiently detailed to provide significant help to actors with the means to cause substantial harm. That said, as the size of these models grow, their capacity to respond to detailed, domain-specific queries will likely increase. This will present a challenge to AI companies releasing frontier models, as they seek to provide models which can help with research, but not provide information which could assist irresponsible or malicious actors. One potential solution would be to provide vetted researchers or labs with adequate safeguards access to versions of models with greater dual-use potential. In addition, executing a complex series of operations like those required for bioterrorism involves additional steps beyond current capabilities \autocite{mouton_operational_2024}. Acquiring the necessary materials for bioterrorism would require navigating real-world physical barriers. Continuing to monitor physical barriers will provide an important layer of protection on top of model safeguards.

Capability uncertainty is another source of decreased distinguishability. In contrast to the clarity of traditional armaments (knowing the location and range of a missile, for example) the advancements in foundation models are less transparent. This inherent uncertainty in the capabilities of foundation models complicates the verification of a state's intentions \autocite{anderljung_frontier_2023}. While general improvements in new models can be anticipated because of scaling laws, advancements within specific domains are difficult to predict. Civilian applications of these models are often openly tested \autocite{huggingface_open_nodate}, in contrast to military applications, which are typically conducted covertly. Both the US and China are known to be exploring military applications of foundation models, yet the specific use cases being developed remain opaque. This lack of visibility into the other's advancements contributes to decreased distinguishability, complicating the assessment of potential military capabilities.\footnote{We lack the space here to look at the importance of opacity in certain contexts like nuclear stability or off-stage signaling. This would require a longer discussion.}

As a result of the limitations in our ability to understand a model directly, capabilities need to be “revealed” through red-teaming, evaluations, prompt engineering, monitoring, and experimentation \autocite{openai_gpt-4_2024}. As I discuss in more detail later, evaluations offer one way of improving distinguishability. Researchers currently use evaluations to understand the model capabilities across a broad range of domains. For instance, evaluations can assess a model’s ability to analyze documents or complete coding tasks. They are also used to test various safety-relevant aspects of a model. Researchers run dangerous capability evaluations to test whether a model can manipulate, deceive, blackmail, or carry out other potentially undesirable behaviors. Evaluations for defense or military foundation models are not widespread, but researchers should develop them so that states can better understand the capabilities of models in a military context.

\subsection{System Use Cases}

\subsubsection{Hardware and Software}

A system encompasses the wider operational context in which a foundation model operates. This includes not only the model itself but also the supporting hardware, data pipelines, user interfaces, security protocols, and other components essential for its functioning. In military contexts, this extends to how the model integrates with military hardware and command-and-control systems. The term “use cases” refers to the specific tasks users intend for the model to perform and how they guide its operation. These use cases, ranging from data analysis and military planning to media generation, play a critical role in defining civilian-military distinguishability. While we do not yet have enough information to develop a comprehensive typology for each potential military use case, I will attempt to outline some key considerations such a typology would include.

Whether a system is civilian or military depends on who the user is. Thus, distinguishability depends on identifying the motivations of a potential adversary actor through context. Since foundation models generally reduce the cost of shifting between civilian and military uses, inferring intentions may be more difficult. The problem of attribution is particularly difficult when interacting with systems in digital environments. Offensive cyber operations, such as hacking or deploying malware, can often be conducted covertly, making it difficult for an observer to identify the threat and the motivations until after an attack has occurred. Penetrating a network could be defensive and then later used for offensive purposes \autocite{lindsay_tipping_2015}. Moreover, in a crisis scenario, de-escalation becomes more difficult if you cannot identify the culprit or lack information on how such a de-escalation could occur. Conversely, if an adversary is responsible in a scenario which requires punishment, lack of confidence in this could undermine deterrence. 

Military hardware, being more specialized than foundation models, offers clearer distinguishability. In scenarios like aerial surveillance, where hardware is involved, intentions may be discernible through observable actions. Hardware type, testing and deployment patterns all provide insight into an actor's intent.

While a model might theoretically predict a missile's path, it may not physically defend against it because of the specialized nature of missile systems, which have limited ranges and substantial repurposing costs. To illustrate this, consider a model which contains knowledge in a given domain, but is weak at generalizing this information in novel contexts. Under the status quo, deployed weapons systems can be used to carry out only a small number of tasks. Defense systems like the Aegis Ballistic Missile Defense System use traditional ML models that can defend against short and medium range missiles but would fare poorly at other tasks, like defending against longer range missiles or conducting offensive operations \autocite{lockheed_martin_artificial_2023}. Deploying Aegis can thus signal the US's intention to defend against specific threats without demonstrating hostility toward other states. For instance, in 2009 Vladamir Putin was supportive of the US deploying Aegis in the Black Sea, since this would not be very effective at defending against Russian missiles \autocite{barnes_russias_2009}. This enabled the US to signal to Russia that it was focused on defending against Iran. This lack of generalizability aided in distinguishing intentions with respect to particular states. Foundation models, on the other hand, are better able to generalize to novel contexts and so a system that can predict the trajectory of medium range weapons may also be able to generalize to longer range weapons.

All this might result in increased difficulty in signaling narrow defensive intentions. But inflicting harm often involves solving a wide set of problems. In the example above, for Aegis to carry out longer distance attacks, it would likely require changes to its military hardware. This would involve multiple steps of development and deployment, which a frontier foundation model is far from being able to carry out today. 

Safety measures, like preventing platform misuse, help minimize the risk of systems being used in a military context. Since all companies producing frontier models are currently American, this provides some guardrails on the types of military contexts in which such models might be used. Open source models, where the model weights are readily accessible and do not contain platform safety measures, enable greater scope for use by militaries. Without such measures, more state and non-state actors can use it and test it for their military platforms. Unlike proprietary models, users can change the weights of open source models, thus undoing safety mitigations from post-training. This lowers the barriers to entry for more actors modifying model capabilities and using these models in military contexts.

\subsubsection{Information Environment}

The dual capacity of foundation models to produce and analyze information has contrasting implications for distinguishability. Foundation models reduce states’ confidence in the information environment, but they also enhance transparency by improving access to information. In terms of the former, I identify two factors affecting the information environment: production and dissemination, and interpretability. In terms of the latter, foundation models will improve data fusion and analysis. Further research is needed to understand if foundation models will, on net, increase the accuracy of information available or decrease it.

Foundation models reduce the cost of producing and disseminating information \autocite{goldstein_generative_2023}. Generating text, images, video, and sounds will continue to decline in cost production, while the fidelity of AI-generated information will continue increasing, making it more difficult to know what is AI generated and what is human generated. Much of this information may be generated to fool the adversary or be created by third-party actors with other intentions. Increased dependence on foundation models for data analysis increases the incentives to alter the information environment through poisoning the training data or conducting adversarial attacks \autocite{geist_deterrence_2023}.

Foundation model outputs are also challenging to interpret. Because of the size and complexity of neural nets, foundation models are largely black boxes, giving us a limited understanding of why any given output is produced. Some techniques enable us to partially understand the reasoning behind an output, such as chain-of-thought and various interpretability approaches \autocite{lanham_measuring_2023, bills_language_2023}. These are making progress, but are still not reliable enough to depend on, particularly in complex and high-stakes cases. Making interpretability tools specific to military AI systems has numerous technical hurdles, given reliability requirements.

Understanding the intentions behind a model's actions forms a critical component of the challenge, essential for establishing trust and averting unintended escalations arising from misinterpreted intentions. Foundation models may exceed human transparency in some respects, given that human behavior lacks complete reasoning transparency. Humans have many social and institutional mechanisms for building trust; the DoD, for instance, is attempting to build trust in smaller models through human-AI teaming in programs like NGAD \autocite{osborn_air_2023}. Though the specifics of this program are not publicly disclosed, human-AI teaming involves the defense operator coordinating with an AI to achieve some task. Typically, this means automating some aspect of the defense operator’s workflow, while enabling the operator to exercise judgment or exert control when necessary \autocite{committee_on_human-system_integration_research_topics_for_the_711th_human_performance_wing_of_the_air_force_research_laboratory_human-ai_2022}. Ensuring that the machine provides appropriate feedback to the human is one of the key factors in building trust.

Foundation models will also have a contrasting effect, improving our analytical capabilities. A key constraint on intelligence value, such as documents, satellite imagery, and sensor data, lies in identifying patterns in the most salient data points. Intelligence and defense analysts are increasingly bottlenecked by a lack of time or resources to identify the most salient data, as the cost of satellites and sensors decline. Foundation models excel at filtering and consolidating vast amounts of data, giving analysts a greater ability to extract meaningful insights more effectively. This bolsters distinguishability by facilitating states in discerning the objectives of their counterparts. Thus, even though the cost of producing false information or spoofing decreases, the effects may be counterbalanced by improved data fusion and analysis from foundation models.

Model capabilities in novel pattern detection will likely continue the trend of increased battlespace transparency. Inferring the intentions of an opponent involves a variety of factors, such as how they are positioning their military assets, what components they are purchasing, and how they are communicating. Naval operations, for instance, typically have low distinguishability because of the lack of information on an adversary at sea and the flexibility of naval weapons \autocite{caverley_cruising_2020}. Combining inexpensive sensors with models running inference on the collected data could increase observability \autocite{glonek_coming_2024}. 

Foundation models have the potential to enhance National Technical Means (NTM) by making it easier for adversaries to verify treaty compliance. During the Cold War, NTM was essential for arms control verification, but the specific methods used were kept secret to protect reconnaissance capabilities \autocite{bateman_trust_2023}. Despite this secrecy, a range of tools --- such as satellites, radar, and sensors for detecting sound waves, seismic activity, and particles --- were likely used \autocite{baker_nuclear_2023}. However, the falling costs of sensors have led to a flood of data, making the task of identifying key data points into a laborious and time-consuming process \autocite{vaynman_better_2021}. Foundation models can play a role in addressing this challenge. They are capable of integrating data from various sensor modalities, which is increasingly important given the growing volume of data \autocite{committee_on_the_review_of_capabilities_for_detection_verification_and_monitoring_of_nuclear_weapons_and_fissile_material_technical_2023}. This may make it more costly to spoof, since convincing spoofing requires the adversary to falsify data across every modality.

The declining cost of data collection, data labeling, fusion, and analysis may lower the barriers to sophisticated data analysis, potentially enabling a broader range of entities (including the private sector and non-profits) to participate in verification processes \autocite{hanham_remote_nodate}. This development might lead to a more distributed and robust verification landscape, but it remains to be seen how significant this shift will be. The potential for AI verification is yet to be fully tested, but researchers and organizations should conduct this testing across a range of domains including monitoring CBRN materials, tracking compute, verifying compliance with export controls and sanctions, and identifying human rights abuses.

\subsection{System Deployment}

The deployment method of foundation models significantly impacts distinguishability, presenting two primary questions. First, there is the degree of specialization between civilian and military applications. Second, there is the level of spillover in capabilities between the civilian and military sectors. These problems will be addressed through domestic regulations and norms, as well as internationally, as states endeavor to develop and incorporate these models into their military systems.

The first key question is whether the models deployed in civilian settings resemble those deployed in military contexts. If the similarity is high, it becomes challenging to distinguish between civilian and military applications. If the models excelling in enterprise contexts are less effective for military purposes, it becomes feasible to differentiate their applications. Military and commercial drones, for instance, use different ML models than those developed by frontier AI companies. The class of problems a UAV is confronted with tends to require more spatial navigation, whereas consumer and enterprise AI assistants require more generalized analytical capabilities. Moreover, while AI assistance will largely continue relying on the cloud for most inference, this may not be suitable for UAVs, as latency would be a bigger problem. On-device inference, on the other hand, would not be as powerful, since the size and weight specifications of a UAV would inherently limit the available compute. But it remains to be seen whether this will hold true for other applications like military planning or weapons development, where latency and design specifications are less of a limiting factor than with UAVs.

Scientific, engineering, and military research are areas where civilian and military models are likely to be similar, presenting significant challenges in maintaining distinguishability. General capabilities in research and analysis are likely to be useful across many sectors. Moreover, the inherently broad nature of basic research makes it particularly difficult to distinguish between military and civilian research. States may also resist imposing limits on R\&D because of the broad potential applications and the critical importance of scientific advances.

A possible strategy to address this challenge is to steer the development of foundation models so their military applications are not as versatile as their economic applications. Given that the current paradigm involves training increasingly general models, establishing limitations on weapons platforms is a more viable solution than establishing limits on the models themselves. This approach would enable states to harness the economic potential of these models while mitigating the risk of military escalation.

Internationally, the growing recognition of the economic benefits of foundation models is likely to lead to an increased interest among states in their development. As the potential military uses of these models become more apparent, influencing the strategic direction of their development will become increasingly important. Major powers will likely develop strategies aimed at controlling the spread of these models, particularly among rival states. 

States may use the threat of developing foundation models as a bargaining chip. Japan, while not actively pursuing a nuclear weapons program, has the capability to produce nuclear weapons, thereby holding strategic leverage in security dialogues. Private companies in the US may play a role in providing US allies with significant civilian and military capabilities which may disincentivize the development of large models outside the US. States may seek to hedge by building smaller models or fine-tuning open source models.

The second question around deployment is the extent to which capabilities in the civilian sector translate into military advances. This relationship can be thought of in terms of a conversion rate from civilian to military capabilities. An increase in model capabilities in the civilian sector (x percent) leading to a comparable increase in military capabilities (y percent) implies a strong diffusion from civilian to military uses. A y/x ratio close to 1 indicates low distinguishability due to strong civilian-military ties. Presently, this ratio is low, meaning the military integration of foundation models is very limited. Returning to an example from earlier, if foundation models begin to drive scientific advancements, their broad nature could lead to significant engineering advances in the military sector. In this scenario, foundation models would become highly important strategic assets. Deploying a foundation model with civilian intentions could present a significant threat to adversary states if they are limited in their access to this model.

The civilian-military conversion rate is partly determined by the high costs and uncertainties associated with deploying models in military contexts, as well as limitations in the capabilities of existing models. Current models lack the robustness and reliability required to operate in many military contexts, making small capability improvements in the civilian sector insufficient for broad military adoption. The automation of tasks also necessitates significant model testing and the training of personnel in new workflows, increasing integration barriers. However, as these barriers decrease, improvements in civilian sector models are likely to increasingly permeate the military sector.

The private industry and its institutional ties to the military also play a role \autocite{brandt_defense_1994}. Many of the current advances in capabilities are coming from privately funded companies that derive their revenue from civilian customers. This contrasts with previous eras, where companies like Fairchild Semiconductor received up to 95 percent of their revenue from the military or NASA \autocite{miller_assault_1971}. In this case, inputs, capabilities, and use cases were more likely to be optimized for a military context, resulting in a higher civilian-military conversion rate.

As discussed earlier, AI labs tend to have a range of safety measures developed during the course of training a model, as well as usage and enforcement policies which seek to prevent misuse. Among other uses, this can diminish the extent to which state and non-state actors can use the models in unauthorized contexts. This strategy may be viable because of the cost structure of frontier foundation models, which typically have high fixed costs for initial training, while inference costs are variable, depending on usage. This high entry barrier means fewer actors are involved in developing these models, simplifying coordination during training and deployment. 

As I will further discuss in the next section, practices such as open red-teaming, rigorous evaluations, and incident sharing have become important for promoting transparency in research. These approaches can help states clarify where the boundaries lie between civilian and military uses. By fostering an environment of openness and collaboration, states can better navigate the ambiguities inherent in foundation models, ensuring that advancements contribute positively to scientific progress while mitigating risks associated with dual-use capabilities.

Regulatory frameworks play a crucial role in shaping the application of foundation models, especially in sectors with stringent regulations, like defense. Establishing clear functional boundaries within the capabilities of these systems can help states achieve economic advancement without intensifying military competition. The extent to which regulations enforce a clear demarcation in the use of foundation models will be crucial in managing their dual-use capabilities effectively. However, since many potential regulations are still unclear, the private sector currently serves as a more flexible testing ground where different approaches to safety can be explored and refined.

\section{Toward Distinguishability}

The GPT dilemma provides a framework for considering the trade-offs in international rules or agreements at each stage of the foundation model deployment cycle. Implicit and explicit security trades have been important in preventing the proliferation of potentially dangerous technology. The Intermediate-Range Nuclear Forces (INF) Treaty is one example, which demonstrates how distinguishability enables states to engage in effective security trades. Using the INF as a case study, this section will explore how improving distinguishability enables states to negotiate viable "security trades." I suggest four approaches for managing the development and deployment of foundation models within a military setting:
\begin{enumerate}
    \item establishing red lines for military competition,
    \item fostering information sharing,
    \item implementing national technical means for verification, and
    \item constraining weapons platforms.
\end{enumerate}

In the context of the INF Treaty, effectiveness hinged on precisely differentiating the capabilities of specific missile systems, particularly prohibiting ground-launched cruise missiles with ranges between 500 to 5,500 km \autocite{jones_reagan-gorbachev_2012}. Although the INF Treaty dissolved in 2019, its role in stabilizing US-USSR relations and enhancing verification and information-sharing practices render it a valuable case study for contemporary considerations of stability. The treaty's approach was also effective in setting narrow red lines for the competitive development of technologies.

The INF Treaty was made possible by the agreement between the US and USSR that intermediate-range weapons were destabilizing and that they were sufficiently confident they could identify their counterparts’ weapons platforms associated with these capabilities. Similarly, for states to agree to restrictions on foundation models, there first needs to be a consensus between the US and China around what common risks foundation models pose. Though no such consensus currently exists, there are several fruitful areas to investigate.

One potential strategy would be to focus on limiting the generalizability of certain model capabilities. Foundation models with highly specialized functions might offer some degree of distinguishability because of their narrow scope. This aspect becomes crucial when differentiating between models used for military purposes and those employed in economic sectors. If military-oriented foundation models distinctly differ from those used for economic activities, clear civilian-military demarcations are made. 

Alternatively, states could establish red lines prohibiting the development of foundation models with certain high-risk functionalities, like self-replication. Targeted restrictions would clearly delineate allowable and prohibited technologies or capabilities. Implementing safety mitigations in frontier models could serve to limit their potential for offensive use. For instance, GPT-4, before safety mitigations were in place, had the capability to generate content that could pose risks, including details related to CBRN hazards \autocite{openai_gpt-4_2024}. Anthropic’s Responsible Scaling Policy \autocite{anthropic_anthropics_2023}, OpenAI’s Preparedness Framework \autocite{openai_preparedness_2023}, and Google’s DeepMind’s Frontier Safety Framework all provide outlines of different risk thresholds which models could reach at different capabilities \autocite{google_deep_mind_frontier_nodate}. They also outline ways of red-teaming models for these threats and mitigation approaches to prevent such models from posing threats. 

While jailbreaks and other countermeasures exist, the safety mitigations in place significantly decrease the chances of a model producing harmful outputs in civilian models, thereby aiding in the clear separation of civilian and military applications. With open source models however, developers have direct access to the model's weights. In such cases, safety mitigations can be easily bypassed, making distinguishability more difficult. 

Even with proprietary foundation models it is difficult to ensure that state actors can verify adherence to red lines. This could be improved by a Digital Treaty on Open Skies. Drawing from the Open Skies Treaty \autocite{britting_open_2002}, which enabled signatory states to conduct unarmed aerial surveillance flights over each other’s territories to collect data on military forces and activities, a Digital Treaty on Open Skies would permit states to mutually inspect each other's foundation model capabilities. This would be achieved through structured evaluations and red-teaming exercises designed to cultivate transparency and trust, but with a focus on foundation models rather than traditional military hardware. Initially, such mechanisms might be adopted among allied states to reduce ambiguities in their interactions and encourage broader trust. As foundation models contribute to an increasingly large share of state military capabilities, it will become more important for states to observe the capabilities of other states’ models. 

Onsite inspections played a pivotal role in the verification process of the INF Treaty, providing a direct method for ensuring compliance with its provisions \autocite{russell_-site_2021}. These inspections allowed for the physical verification of treaty obligations by enabling inspectors to count missile units, examine serial numbers, and directly observe the destruction of missile systems. One could envision a similar approach for AI where states inspect each other's data center operations. However, this concept would require further development before we could assess its feasibility. There would also be considerable challenges if we sought to inspect military hardware. While onsite inspections are highly advantageous for verifying missile capabilities, it would be difficult to confirm the use of a specific model in a weapons platform. This difficulty is compounded by the ease with which software can be modified, making it challenging to ensure that a particular model is being used consistently in a given piece of military equipment. 

Currently, frontier models are primarily assessed with behavior-based verification methods, such as evaluations and red-teaming exercises. These existing efforts should be expanded, while also investing in verification through remote sensing and analysis. Evaluations involve quantitatively assessing a model's proficiency in executing specific tasks, such as conducting cyberattacks or self-replication, without disclosing the sensitive details of its development. Given that evaluations represent a relatively new assessment methodology, the consistency of their quality can significantly vary. Current evaluations also have a range of problems, such as the lack of consistency between how evaluations are conducted and an inability to verify whether the questions posed in an evaluation are in the training data. The nature of evaluations, which test models against a predefined set of scenarios, raises concerns that exclusive reliance on them may miss unidentified risk factors. It is essential to complement evaluations with red-teaming exercises, which are designed to uncover hidden vulnerabilities and risk vectors that evaluations alone might not detect. This combined approach ensures a more comprehensive assessment of a model's capabilities and potential risks. 

An important aspect of inspections working under the INF was that inspectors could obtain adequate information for verification while minimizing the perceived risk by either side that this would reveal additional state secrets \autocite{vaynman_better_2021}. Since anyone can run evaluations on publicly available models, running evaluations and sharing the results would generally not risk compromising intellectual property or disclosing sensitive intelligence.\footnote{Unless the evaluations highlighted some catastrophic vulnerability.} Moreover, sharing the outcomes of evaluations and red-teaming exercises is often beneficial, as it provides insights into their likely behavior and aids in building trust by highlighting their capability gaps. 

Running evaluations on unreleased models or models to be used in a military context presents a more challenging problem. States will want to fine-tune models on sensitive data like reconnaissance satellite images and develop their own evaluations to assess the capabilities of foundation models for their particular needs. But as with the INF, how and with whom this information is exchanged may be sensitive. In the context of foundation models, leaders might exhibit caution in sharing evaluation results, particularly when such disclosures could pose security risks. For example, a state might withhold information if an evaluation reveals that its model can detect movements of a mobile nuclear launcher.\footnote{While there are inherent limitations to the level of transparency that might be desired – such as concerns over compromising second-strike nuclear capabilities – the actual risk of such “use it or lose it” scenarios is low for now; see \autocite{geist_deterrence_2023}.} 

For more sensitive topics like national security and CBRN, information-sharing protocols are currently being developed between AI labs in the private sector \autocite{openai_frontier_2023}. These also include a component of sharing pertinent information with the White House \autocite{the_white_house_fact_2023}. Interstate initiatives for information sharing are likely to start among allied states, aiming to improve interoperability between their militaries and greater transparency on emerging capabilities. Sharing evaluations between the US and China would have to start in less sensitive domains, but enhance trust for future agreements. Thus, carefully managed information sharing and civilian experimentation are important components in improving distinguishability.

National technical means (NTM), integral to the verification process of the INF Treaty, involved using satellite imagery and other data to monitor compliance. Similarly, foundation models can enhance distinguishability through advanced technical verification measures. The integration of closed and open source intelligence with low-earth orbit satellites and inexpensive sensors can be significantly augmented by foundation models. As discussed earlier in the paper, this is likely the most promising area for improving distinguishability, though insufficient research exists exploring this possibility.

To develop more effective verification methods, domain experts should evaluate the capabilities of foundation models, focusing on adherence to red lines and compliance with treaties, export controls, and sanctions. Initial efforts should begin with researchers creating evaluations and conducting red-teaming exercises for tasks such as data labeling and change detection. These efforts can then expand to experiments and test beds to assess the viability of foundation models in real-world conditions for intelligence and military analysts. As model capabilities improve, these methods can be applied to more complex tasks, such as analyzing weapon platforms’ deployment patterns to gain insights into a state’s intentions. While human analysts may struggle with the subtleties of these patterns, future foundation models, fine-tuned on relevant data, could potentially identify novel patterns.

Leaders in the private sector can contribute by developing technologies and collaborating with organizations to enhance data fusion and analysis through the use of foundation models. Implementing systems for models to effectively "identify themselves," akin to physical markers in conventional arms control, faces the challenge of digital identity manipulation, underlining the need for robust verification mechanisms. But the ability to verify a state's behavior across various domains—such as imagery, infrared, text, and radio—complicates deception efforts, enhancing verification efficacy.

States can provide reassurance in the realm of new technologies by focusing on areas with clearer distinguishability. To achieve this, states could negotiate agreements on model deployment where distinguishability is more evident, or engage in strategic bargaining, trading advantages in less distinguishable domains for concessions in areas where they lag but the capabilities are clearer \autocite{snyder_limiting_1988}. Following this approach, distinguishability may be possible through placing limitations on how foundation models are combined with military systems and delivery platforms. Currently, foundation models primarily conduct digital tasks, but could be combined with physical systems like AWS, missiles, and other weapons platforms. States could place limitations on delivery platforms or update arms control agreements to account for capability improvements resulting from foundation model integration. This would be more verifiable, but would still have significant challenges given the current international environment.

\section{Conclusion}

This paper offers several novel contributions. It lays out a framework for thinking about the role of foundation models in a dual-use context, exploring the interplay between these models and distinguishability in international relations. It highlights the growing importance of foundation models as general-purpose technologies and their potential to influence deterrence dynamics because of their dual-use capabilities. Through a detailed examination of the four factors in the foundation model development cycle, the paper demonstrates how the inputs, model capabilities, system use cases, and deployment of these models can impact the ability to discern between civilian and military intentions. As foundation models continue to advance and integrate into military operations, understanding their implications becomes imperative.

This paper also provides a means of assessing the trade-offs for policymakers and AI labs. IR scholarship needs to grapple with the importance of foundation models for deterrence and conduct further research. AI labs producing foundation models need to be cognizant of the potential repercussions of their innovations, especially considering the increased generalizability of foundation models. Understanding the wider implications can help guide responsible development of these foundation models.

This paper builds on existing scholarship in confidence-building measures and arms control by suggesting several approaches to improve distinguishability between civilian and military uses of foundation models \autocite{shoker_confidence-building_2023, schornig_artificial_2022}. Drawing lessons from the INF Treaty, the paper proposes establishing red lines for military competition, fostering information sharing, constraining weapons platforms, and implementing verification. These measures aim to create clear boundaries, foster trust, and enhance international transparency, addressing the challenges posed by the integration of AI into military operations and contributing to stability in the context of increased great power competition.

\newpage
\clearpage
\printbibliography

\section*{Acknowledgements}

I would like to thank Sarah Shoker, Gretchen Krueger, Cullen O’Keefe, Miles Brundage, Tristan Volpe, Allison Carnegie, Michael Kolhede, Jonathan Reiber, Katrina Mulligan, Andrew Reddie, and Cameron Raymond for their invaluable feedback on drafts of this work. Their insights and suggestions have greatly improved the quality and depth of this work. I am particularly grateful to Miles Brundage for his suggestion of Digital Open Skies.

\end{document}